
---------------------------------------------------------------------------

\input phyzzx
\mathsurround=2pt
\openup 2pt
\hfuzz 30pt
\pubnum{UTS-DFT-92-15}
\pubtype{HE}
\titlepage
\singlespace
\def\a{anomaly}
\def\as{anomalies}
\def\d{\partial}

\def\sfs{super-fields}
\def\sf{super-field}
\def\l{Lorentz}
\def\ieff{I_{\rm eff}}
\def\tj{\tilde J}

\title{General treatment of anomalies in $(1,0)$ and $(1,1)$ two-dimensional
super-gravity}
\author{A. Smailagic\foot{E-mail address: ANAIS@ITSICTP.BITNET}}
\address{International Center for Theoretical Physics, Trieste, Italy\break
Strada Costiera 11-34014, Trieste, Italy}
\andauthor{E.Spallucci\foot{E-mail address: SPALLUCCI@TRIESTE.INFN.IT}}
\address{ Dipartimento di Fisica Teorica\break Universit\`a di
Trieste\break
I.N.F.N., Sezione di Trieste\break
Trieste, Italy}
\vfill
\centerline{In print on: {\it Classical} \& {\it Quantum Gravity} }
\medskip
\centerline{P.A.C.S. : 04.65}
\vfill
\eject
\doublespace
\abstract

In this paper we discuss the interplay among (~super-~)coordinate, Weyl
and \l\ anomaly both in chiral and non-chiral super-gravity
represented by $(1,0)$ and $(1,1)$ two-dimensional models.
It is shown that for this purpose two regularization dependent parameters
are needed in the effective action.
We discuss in {\it full
generality} the regularization ambiguities of the induced effective
action and recover the corresponding general form of the
anomalous Ward Identities. Finally, we explain the difference between
chiral and non-chiral super-gravity models in terms of the free parameters
and establish relation between these two models by projecting $(1,1)$
into $(1,0)$ super-symmetry.

\vfill\eject
\REF\Aa{%
R.Jackiw, R.Rajaraman, Phys.Rev.Lett.{\bf 54}, (1985) 1219;\hfill\break
K.Li, Phys.Rev.{\bf D34}, (1986) 2292;\hfill\break
L.Alvarez-Gaum\'e, E.Witten, Nucl.Phys.{\bf B234}, (1983) 269
}%
\REF\Ab{%
H.Leutwyler, S.Mallik, Z.Phys.{\bf C-33}, (1986), 205;\hfill\break
A.Smailagic, E.Spallucci ``~Chiral Models in Gravity~'', in Proceedings
of the ``~XVII International Colloquium on Group Theoretical Methods
in Physics~'', ed. World Sci. 1988
}%
\REF\Ac{%
For a general review of this topic see:\hfill\break
T.Berger, {\it Fermions in Two (1+1)-Dimensional Anomalous Gauge Theories:
the Chiral Schwinger Model and the Chiral Quantum Gravity}, DESY 90-084,
July 1990
}%
\REF\Aw{%
A.Smailagic, R.E.Gamboa-Saravi, Phys.Lett.{\bf B192}, (1987), 145
}%
\REF\Anoi{%
A.Smailagic, Phys.Lett.{\bf B205}, (1988), 483\hfill\break
A.Smailagic, E.Spallucci, Phys.Lett.{\bf B284}, (1992), 17
}%
\REF\tseyt{%
A.A.Tseytlin, Int.J.Mod.Phys.{\bf A5}, (1990) 1833
}%
\REF\pol{%
A.M.Polyakov, Phys.Lett.{\bf B103}, (1981), 207;
Phys.Lett.{\bf B103}, (1981), 211
}%
\REF\Ae{%
M.B.Green, J.H.Schwartz, E.Witten, ``~super-string Theory~'', vol.1, Cambridge
Univ. Press 1987
}%
\REF\Ad{%
S.J.Gates Jr., M.T.Grisaru, L.Mezincescu and P.K.Townsend, Nucl.Phys.{\bf
B286},
(1986), 1
}%
\REF\evans{%
M.Evans, B.Ovrut, Phys.Lett. {\bf B175}, (1986), 145
}%
\REF\fre{%
P.Fr\'e, F.Gliozzi, Nucl.Phys.{\bf B326}, (1989), 411
}%
\REF\Af{%
M.T.Grisaru, L.Mezincescu and P.K.Townsend, Phys.Lett.{\bf B179}, (1986), 247
}%
\REF\leut{%
H.Leutwyler, Phys.Lett. {\bf B153}, (1985), 65
}%
\REF\Ag{%
S.J.Gates, H.Nishino, Class.Quantum Grav.{\bf 3}, (1986), 391;\hfill\break
R.Brooks, S.J.Gates Jr., Class.Quantum Grav.{\bf 5}, (1988), 367
}%
\REF\gris{%
M.T.Grisaru, R.M.Xu, Phys.Lett{\bf B205}, (1988), 486
}%
\REF\Ah{%
A.Smailagic, Phys.Lett.{\bf B196}, (1987), 499
}%
\REF\Ai{%
R.Brooks, F.Mohammad, S.J.Gates Jr., Class.Quantum Grav.{\bf 5}, (1988), 785
}%
\REF\Ajose{%
A.Smailagic, J.A.Helayel-Neto, Mod. Phys. Lett. {\bf A10}, (1987), 787
}%
\refsend

\chapter{Introduction}

Anomalies are one of the central topic in modern quantum field theories and
can be considered from a variety of perspectives ranging from the construction
of quantum consistent phenomenological models of particle interactions to
geometrical formulation of gauge theory.

One of the most intriguing  aspect of the problem is the similarity
between the anomaly pattern in two-dimensional abelian gauge theories and
bi-dimensional fermionic quantum gravity [\Aa,\Ab,\Ac].
The Schwinger model quantum ambiguities
can be conveniently described in the effective action by a
single free parameter interpolating between vector and axial
symmetry preserving regularization schemes, and shifting \a\ from the
gauge to the axial current [\Aw].
In fermionic quantum gravity gauge and axial symmetries are
replaced by Lorentz and Weyl invariance, so it is tempting to transfer the
above scheme to the gravitational case as well. However, a third local
symmetry,i.e. general covariance, has to be taken into account requiring
a more general formulation of the problem.
Just recently we proposed a general regularization scheme where all the
three gravitational symmetries (Lorentz, Weyl, and general coordinate
transformations) are treated on the same footing at the quantum level
[\Anoi].
The formal analogy with the Schwinger model can be recovered under
special conditions showing that the gravitational models contains
more information due to more symmetries.

Renewed interest in 2-dimensional induced super-gravity, mainly in
connection with strings theories and super-Liouville models
[\tseyt] , makes it
important to address the problem of anomalies in this case [\pol].
In general
chiral and non-chiral models show a different anomalous behavior
in the sense that the former always posses anomalous
Ward Identities [\Aa], while the latter allows the
shifting of \as\ from one current to another, by properly
choosing the regularization scheme. In order to discuss
both physical situations we shall consider $(1,0)$ super-gravity as an example
of chiral model, and $(1,1)$ super-gravity as a non-chiral model. In ordinary
gravity it is possible to consider both cases on the same
footing [\Anoi], while in the super-symmetric case such a
simplified approach is not possible because, as shown later, projection
of higher-N supersymmetric model to a lower one is a complex procedure
that involves projections of covariant derivatives and cannot be described
by simple use of a generalized chiral projector
$P_\beta={1\over 2}(1+\beta\gamma^5)$. Therefore,
we shall separately consider the two models. The paper is planned as follows.

In Sect.2
we shall construct the most general form of the induced effective
action for $(1,0)$ super-gravity in terms of arbitrarily weighted
local counterterms, and study the interplay among various quantum symmetries.
Though we start with a number of arbitrary parameters the result boils down
to two parameters necessary to describe super-gravitational \as\ in their
full generality, as long as the super-connection is the main ingredient in
constructing the effective action.

In Sect.3 we shall establish connection between chiral and non-chiral
super-gravity models by projecting  $(1,1)$ \sfs\ in terms
of $(1,0)$ \sfs.

Sect.4 is devoted to a brief summary of the results.

\chapter{ $(1,0)$ Induced super-gravity}

Superfield formalism for $(1,0)$ supergravity is already known [\Ad],
[\evans],[\fre],
and is described in terms of unconstrained prepotential
\sfs\ some of which are gauge degrees of freedom.
These redundant (gauge) degrees of freedom can be removed by an
algebraic gauge choice $H_L^+=0$, thus constraining
parameter \sfs\  $K^L$ of super-symmetry transformation.
After this procedure one is left with four
{\it unconstrained prepotential \sfs\ } $H_L^{(1,0)-},\,H_-^{(1,0)+},\,
H_L^{(1,0)L}$ and
$H_-^{(1,0)+}$ relevant for the symmetries whose \as\ we want to investigate.
Their {\it linearized} super-symmetry transformations are\foot{
We shall use the following notation : + and -  refer to light-cone components
of bosonic quantities, while L and R labels left and right chirality components
of fermionic quantities. Interior product between bosons is defined as
$$
A_aB^b={1\over2}(A_+A_- +A_-B_+)\ .
$$
and for fermions is
$$
\psi^\alpha \psi_\alpha = 2i\psi_L\psi_R
$$
The two kind of indices are related by $LL\equiv +$, and $RR\equiv -$.

Finally, our notation can be translated into the language of ref.(4) according
to the substitutions
$$\eqalign{
&L\rightarrow +\cr
&R\rightarrow -\cr
&-\rightarrow =\cr
&+\rightarrow \ne\cr}
$$
}
$$\eqalign{
&\delta H_L^{(1,0)-}=-D_L K^-\ ,\qquad
\delta S^{(1,0)}={1\over 2}(\d_- K^- +\d_+K^+)+\Lambda
\cr
&\delta H_-^{(1,0)+}=-\d_- K^+\ ,\qquad
\delta L^{(1,0)}={1\over 2}(\d_- K^- -\d_+K^+)+K
\cr}
\eqno(2.1)
$$
where $K^+,\,K^-,\,K$ and $\Lambda$ are coordinate, \l\ and Weyl parameter
\sfs. $S^{(1,0)}$ and $L^{(1,0)}$ are Weyl and \l\ \sf\ compensators defined,
at the linearized level, as
$$\eqalign{
&S^{(1,0)}=-\left(H_L^{(1,0)L}+{1\over2}H_-^{(1,0)-}\right)\cr
&L^{(1,0)}=\phantom{-}\left(H_L^{(1,0)L}-{1\over2}H_-^{(1,0)-}\right)\cr}
\eqno(2.2)
$$
and $D_L={\partial\over\partial\theta^L}+i\theta^L\partial_+$
is the ordinary super-symmetric derivative .

Variation of a generic action with respect to independent prepotentials
gives
$$\eqalign{
\delta I=\int d^2x\, d\theta^L \Bigl(&\delta H_L^{(1,0)-}J^{(1,0)}_{--}+
\delta H_-^{(1,0)+}
J^{(1,0)}_{+L}+\cr
&\delta H_L^{(1,0)L} J^{(1,0)}_R+\delta H_-^{(1,0)-}
\tj^{(1,0)}_R\Bigr)
\cr}
\eqno(2.3)
$$
where $J^{(1,0)}_{--},\,J^{(1,0)}_{+L},\,J^{(1,0)}_R$ and $\tj^{(1,0)}_R$
are the corresponding
super-currents. If the action (2.3)
is required to be invariant under linearized
super-symmetry transformations $(2.1)$, then one finds the {\it classical}
conservation laws under:

\noindent
i) general-super-coordinate invariance
$$\eqalign{
&\d_-J^{(1,0)}_{+L}+{1\over 2}\d_+J^{(1,0)}_R=0\cr
&D_L J^{(1,0)}_{--}+\d_-\tj^{(1,0)}_R=0\ ;\cr}
\eqno(2.4)
$$
ii) super-Weyl invariance
$$
\tj^{(1,0)}_R+{1\over2}J^{(1,0)}_R=0\ ;
\eqno(2.5)
$$
iii) super-\l\ invariance
$$
\tj^{(1,0)}_R-{1\over2}J^{(1,0)}_R=0\ .
\eqno(2.6)
$$
Whenever \l\ invariance is {\it assumed} to be preserved at the quantum
level, one can further gauge away \l\  compensator $L^{(1,0)}$ thus
constraining
the parameter \sf\ $K$. However, regularization of
induced effective action in quantum field theories is an intrinsically
ambiguous procedure and
there is no {\it a priori} reason to privilege one symmetry among the
others. Regularization ambiguities manifest themselves in the
effective action of the anomalous models as arbitrary parameters interpolating
among different regularization schemes, preserving one or the other, or
none of the classical symmetries [\Aa,\Aw,\Ac].
With this remark in mind we construct
the {\it general form of the effective action} for $(1,0)$ induced
super-gravity:
$$\eqalign{
\ieff={1\over 96\pi}\int d^2x\, d\theta^L
\Bigl[&{\bf A} D_L H_L^{(1,0)-}\,{1\over\nabla^2}\,\d_-^4H_L^{(1,0)-} +
{\bf B} D_L H_-^{(1,0)+}\,{1\over\nabla^2}\,\d_+^3H_-^{(1,0)+} +\cr
&c H_L^{(1,0)-}\,\nabla^2\, H_-^{(1,0)+} +dS^{(1,0)}\d_-^2 H_L^{(1,0)-}
+eS^{(1,0)}\d_+D_L H_-^{(1,0)+}\cr
+&fS^{(1,0)}\d_-D_L S^{(1,0)}
+g L^{(1,0)}\d_-^2H_L^{(1,0)-} +mL^{(1,0)}\d_+D_L H_-^{(1,0)+} \cr
+&n L^{(1,0)}\d_-D_L L^{(1,0)} +vL^{(1,0)}\d_-D_L S^{(1,0)}\Bigr]\cr}
\eqno(2.7)
$$
where $\nabla^2$ is the covariant D'Alembertian, and the coefficients
{\bf A} and {\bf B} of the {\it non-local} part of $(2.7)$ are
{\it uniquely} fixed by the contributions from the matter part of the
classical action.\foot{From now on, we shall suppress the global numerical
factor $1/96\pi$ in front of the effective action.}
These numbers have been perturbatively computed, for instance, in
super-string theories [\Ae],[\Af]. All the other
coefficients are regularization dependent and can be fixed according to
which symmetry one wishes to preserve at the quantum level (hopefully
all of the classical symmetries).

{}From the effective action $(7)$ and the symmetry transformations
$(1)$, the following possible anomalies for the individual symmetries
can be derived:
$$\eqalign{
\d_-J^{(1,0)}_{+L}+{1\over 2}\d_+J^{(1,0)}_R=&
\left({m-e\over2}-2{\bf B}\right)\d_+^2D_L
H_-^{(1,0)+}
+\left({g-d\over2}+c\right)\d_-\,\nabla^2\, H_L^{(1,0)-}\cr
+&\left({v\over2}+e-f\right)
\,\nabla^2\, D_L S +\left(m+n-{v\over2}\right)\,\nabla^2\, D_L L\cr
D_L J^{(1,0)}_{--}+\d_-\tj^{(1,0)}_R=&
\left(2i{\bf A}-{g+d\over2}\right)\d_-^3 H_L^{(1,0)-} +
\left(c-{m+e\over2}\right)\,\nabla^2\, D_LH_-^{(1,0)+}\cr
+&\left(d-f-{v\over2}\right)\d_-^2D_L S +
\left(g-n-{v\over2}\right)\d_-^2D_L L\cr
\tj^{(1,0)}_R-{1\over2}J^{(1,0)}_R=&
g\d_-^2 H_L^{(1,0)-} +m\d_+D_L H_-^{(1,0)+} +2n\d_-D_L L
+v\d_-D_L S\cr
\tj^{(1,0)}_R+{1\over2}J^{(1,0)}_R=&d\d_-^2 H_L^{(1,0)-} +
e\d_+D_L H_-^{(1,0)+} +2f\d_-D_L S +
v\d_-D_L L\ .\cr}
\eqno(2.8)
$$
These general results can be simplified by reducing a number of free
parameters through symmetry requirements.
Our request is that the
arbitrary coefficients should be chosen in a way compatible with the
{\it maximum symmetry}
one can obtain at the quantum level. As a first step we require
super-coordinate quantum invariance which fixes parameters in the following
way
$$
\eqalign{
&d=c+2i{\bf A}\phantom{-{\bf B}}\quad m=c+2{\bf B}\quad v=2({\bf B}+i{\bf A})
\cr
&e=c-2{\bf B}\phantom{-{\bf B}\>}\quad g=2i{\bf A}-c\cr
&f=c+i{\bf A}-{\bf B}\quad n=i{\bf A}-{\bf B}-c\ , \cr}
\eqno(2.9)
$$
but induces Weyl and \l\ anomalies
$$
\eqalign{
&\tj^{(1,0)}_R+{1\over2}J^{(1,0)}_R=
(c-2{\bf B})\Sigma^{(1,0)L} -2({\bf B}+i{\bf A})\d_-\omega_L\cr
&\tj^{(1,0)}_R-{1\over2}J^{(1,0)}_R=
(c+2{\bf B})\Sigma^{(1,0)L} +2(c+{\bf B}-i{\bf A})\d_-\omega_L\cr
}
\eqno(2.10)
$$
with \l\ connection super-fields given by
$$\eqalign{
&\omega_L^{(1,0)}=-D_L(S^{(1,0)}+L^{(1,0)})-\d_-H_L^{(1,0)-}\cr
&\omega_-^{(1,0)}=\d_-(S^{(1,0)}-L^{(1,0)})+\d_+H_-^{(1,0)+}\cr
&\omega_+^{(1,0)}=-iD_L\omega_L^{(1,0)}\cr}
\eqno(2.11)
$$
and
$$
\Sigma^{(1,0)L}=(D_L\omega_-^{(1,0)} -\d_-\omega_L^{(1,0)})\ .
\eqno(2.12)
$$
The corresponding effective action is found, from $(2.7)$ and $(2.9)$, to be
$$\eqalign{
\ieff^{(1,0)}=
\int d^2x\, d\theta^L\Bigl[&{\bf A}\Sigma^{(1,0)L}\,{1\over\nabla^2}\,
D_L\Sigma^{(1,0)L}+
({\bf B}+i{\bf A})\omega_-^{(1,0)}\,{1\over\nabla^2}\,\d_+D_L\omega_-^{(1,0)}
\cr
&+(c-2i{\bf A})\omega_L^{(1,0)}\omega_-^{(1,0)}\Bigr]
\ .\cr}
\eqno(2.13)
$$
Notice the appearance of arbitrary coefficient $c$ which is let free by the
requirement of super-coordinate quantum invariance,
and can be further used to eliminate
a local piece of the effective action $(2.13)$. Imposing the above symmetry
requirements have drastically reduced the number of free parameters.

It is suggestive from (2.13) that the construction of the effective action
in terms of super-vierbeins was not the best choice since it resulted in a
number of arbitrary parameters. It is better to work with the spin connection
[\leut]
(2.11) that leave only one parameter in the effective action (~when
coordinate invariance is assumed~). However,
in this case \l\ and Weyl \as\ are detached from the gravitational \a, in the
sense that the free parameter influences only the former. We
would like to present a {\it unique treatment of all three \as .}
To do so one has to design a way of breaking simultaneously general covariance
and, as shown above, a single parameter will not suffice. Since all the
super-vierbeins transform under general coordinate transformation, contrary
to Weyl and \l\ symmetry, presence of all of them in a covariant expression
is necessary to guarantee covariance. Therefore, absence of at least one of
them will spoil general covariance. We shall introduce a second parameter
$b$ in front of some super-vierbeins in a way to obtain generalized
form of the super-connections as
$$\eqalign{
&\bar\omega_L^{(1,0)}=-bD_L(S^{(1,0)}+L^{(1,0)})-\d_-H_L^{(1,0)-}\cr
&\bar\omega_-^{(1,0)}=b\d_-(S^{(1,0)}-L^{(1,0)})+\d_+H_-^{(1,0)+}\cr
&\bar\omega_+^{(1,0)}=-iD_L\bar\omega_L^{(1,0)}\cr}
\eqno(2.14)
$$
with their super-transformations given by
$$\eqalign{
\delta\bar\omega_L^{(1,0)}&=(1-b)\partial_-D_L K^- -bD_L(\Lambda+K)\cr
\delta\bar\omega_-^{(1,0)}&=-(1-b)\,\nabla^2\, K^+ +b\partial_-(\Lambda-K)\cr
\delta\bar\Sigma^{(1,0)L}&=-(1-b)\partial_-D_L(\partial_+K^+ +\partial_-K^-)+
2b\partial_-D_L\Lambda\ .\cr}
\eqno(2.15)
$$
{}From (2.15) one can see that the choice $b=1$ restores general covariance
(linearized super-connections do not transform under general coordinate
transformations) while $b\ne 1$ spoils it.

Starting from the effective action (2.13), where one substitutes $\omega$
with $\bar\omega$ thus introducing dependence on the second parameter, one
can find conservation laws for the symmetries (2.4-6) in a generalized forms
as
$$\eqalign{
\partial_-J_L^{(1,0)} +{1\over 2}\partial_+J_R^{(1,0)}&=
(1-b)\left[-2{\bf B}\partial_+\bar
\Sigma^{(1,0)L} -(c+2{\bf B})\,\nabla^2\,\bar\omega_L^{(1,0)}\right]\cr
D_L J_{--}^{(1,0)} +\partial_-\tilde J_R^{(1,0)}&=
(1-b)\left[c\partial_-\bar
\Sigma^{(1,0)L} +(c-2i{\bf A})\partial_-^2\bar\omega_L^{(1,0)}\right]\cr
{1\over 2}J_R^{(1,0)}+\tilde J_R^{(1,0)}&=
b\left[(c-2{\bf B})\bar\Sigma^{(1,0)L} -
2({\bf B}+i{\bf A})\partial_-\bar\omega_L^{(1,0)}\right]\cr
{1\over 2}J_R^{(1,0)}-\tilde J_R^{(1,0)}&=
-b\left[(c+2{\bf B})\bar\Sigma^{(1,0)L} +
2(c+{\bf B}-i{\bf A})\partial_-\bar\omega_L^{(1,0)}\right]
\cr}
\eqno(2.16)
$$
In this way we set up a unique treatment of all quantum symmetries present in
the $(1,0)$ model. Two independent parameters are needed to achieve
this goal. Assignment of the parameter $b$ interpolates between
super-gravitational
and super-\l\ and Weyl \as. Parameter $c$ has no role in doing that, and
it can only be used to determine the form of the \l\ and Weyl \a\ but
not to remove them. The choice
$b=1$ gives previous results described in (2.10-13), and in this case we
are still free to choose the value of the second parameter $c$ as to
eliminate local piece in (2.13). However, the super-connection
dependence (~and therefore \l\ non-invariance~) persists in the non-local
piece {\it unless} ${\bf A}=i{\bf B}$.
Absence of the \l\ anomaly can be
satisfied only if such condition is fulfilled [\Ad,\Ae].
This is to be expected since $(1,0)$ super-gravity is actually a chiral
model where \l\ \as\  are usually present. Therefore, the above mentioned
relation between coefficients of the non-local piece is not a priori
guaranteed and it depends on the matter contribution to
the effective action . Matter coupling to $(1,0)$
super-gravity is given by the action [\Ad]
$$
I_{\rm matt.}^{(1,0)}=\int d^2x\, d\theta^L (E^{(1,0)}){}^{-1}
\nabla_L\Phi^{(1,0)}\nabla_-\Phi^{(1,0)}
\eqno(2.17)
$$
where $\Phi^{(1,0)}$ is a scalar matter \sf, $(E^{(1,0)}){}^{-1}$
is a super-determinant and
$\nabla_L$, $\nabla_-$ are covariant derivatives. Linearized couplings of the
matter to prepotential \sfs, relevant for calculating non-local pieces of the
effective action, are determined from $(2.16)$ as
$$
\eqalign{
&H_L^{(1,0)-}\left(\d_-\Phi^{(1,0)}\right)^2\cr
&H_-^{(1,0)+}D_L\Phi^{(1,0)}\d_+\Phi^{(1,0)}\ .\cr}
\eqno(2.18)
$$
Now, it is possible to show, by projecting $(2.18)$ in components, that the
non-local part of the effective action in $H_-^{(0,1)+}$ receives both
contributions
from the scalar and fermionic components of the matter \sf, while the one
in $H_L^{(1,0)-}$ receives contribution only from the scalar component. This
produces different coefficients for various non-local terms
$({\bf A}=1\ ,\, i{\bf B}=1+1/2)$. This situation is slightly different
from the non-super-symmetric case where only fermion contribution are
considered, and where ${\bf A}=0\ ,\, i{\bf B}=1/2$. The difference is due to
super-symmetry that introduces super-partners. What is however surprising is
that ${\bf A}$ receives contribution only from the scalar component of the
\sf\ due to the type of coupling.
As a result one has \l\ (and Weyl) anomaly in this model.
As conjectured, this was to be
expected since we are considering a {\it chiral} super-symmetric theory.

Alternatively,we could have eliminated \l\ and Weyl \as\ through a different
choice of parameter $b=0$, thus introducing a gravitational
anomaly. Again, the parameter $c$ can only influence the actual form of the
super-gravitational \a\ but cannot eliminate it.

\chapter{$(1,1)$ Induced super-gravity}

To gain insight about the difference between chiral and non-chiral
super-gravity
models we further consider $(1,1)$ super-gravity as a non-chiral model.
In this case [\Ag] independent
prepotential \sfs\ are $H_L^{(1,1)-},\,H_R^{(1,1)+},\,H_-^{(1,1)-}$ and
$H_+^{(1,1)+}$, whose
symmetry variations are quite similar to $(2.1)$ with appropriate chirality
adjustments (i.e. adding R-chirality pieces).

The $(1,1)$ super-gravity effective action, can be constructed on the
basis of (2.13) and turns out to be [\Ah],[\gris]
$$
\ieff^{(1,1)}=
\int d^2x\, d\theta^L d\theta^R \left[{\bf A}\bar R^{(1,1)}\,{1\over\nabla^2}
\,D_L D_R \bar R^{(1,1)}+
(c-2i{\bf A})\bar\omega_L^{(1,1)}\bar\omega_R^{(1,1)}\right]
\eqno(3.1)
$$
where $R$ is the Ricci scalar \sf\ given in terms of prepotential \sfs\ as
$$\eqalign{
&\bar
R^{(1,1)}=i\left(D_L\bar\omega^{(1,1)}_R+D_R\bar\omega^{(1,1)}_L\right)\cr
&\bar\omega^{(1,1)}_R=bD_R\left(S-L\right)+\partial_+H_R^{(1,1)+}\cr
&\bar\omega^{(1,1)}_L=-bD_L\left(S+L\right)-\partial_-H_L^{(1,1)-}\cr
&\bar\omega^{(1,1)}_+=-iD_L\bar\omega^{(1,1)}_L\cr
&\bar\omega^{(1,1)}_-=-iD_R\bar\omega^{(1,1)}_R\ .\cr}
\eqno(3.2)
$$
We have again introduced two arbitrary parameters in order to treat \as\ in
full generality. In order to extract conservation laws from (3.1) the following
variation is needed
$$
\delta\bar R^{(1,1)}=i(b-1)D_LD_R(\partial_+K^+ +\partial_-K^-)+
2ib\,\nabla^2\,\Lambda
\eqno(3.3)
$$
and we find \a\ relations
$$\eqalign{
D_RJ_{+L}^{(1,1)}+{1\over 2}\partial_+J^{(1,1)}&=
(1-b)\left[2i{\bf A}\partial_+\bar
R^{(1,1)} +i(c-2i{\bf A})\partial_+D_R\bar\omega_L^{(1,1)}\right]\cr
D_L J_{-R}^{(1,1)} +\partial_-\tilde J^{(1,1)}&=
(1-b)\left[2i{\bf A}\partial_-\bar
R^{(1,1)} -i(c-2i{\bf A})\partial_-D_L\bar\omega_R^{(1,1)}\right]\cr
{1\over 2}J^{(1,1)}+\tilde J^{(1,1)}&=
-b\left[(c+2i{\bf A})\bar R^{(1,1)}\right] \cr
{1\over 2}J^{(1,1)}-\tilde J^{(1,1)}&=
b(c-2i{\bf A})\left[\bar R^{(1,1)} -
2iD_R\bar\omega_L^{(1,1)}\right]
\cr}
\eqno(3.4)
$$
As long as the parameter $b$ is concerned the discussion from $(1,0)$
super-gravity repeats itself in $(1,1)$ as well. However, the role of the
parameter $c$ is slightly different. It is still there to further shift between
\l\ and Weyl \a. But, one can have more symmetry in the $(1,1)$
than in $(1,0)$ super-gravity, due to the absence of the non-local
piece in (2.13) in terms of super-connection $({\bf B}=-i{\bf A}; {\rm see}
\quad (3.13))$. So, the parameter $c$ can be chosen
in such a way to eliminate either \l\ or Weyl \a.
This has led to the over-simplified popular belief that, in non-chiral
models \l\ \a\ is always absent, and one can shift the residual
\a\ from the trace
to the divergence of the energy-momentum tensor, and viceversa.
Our result shows that: it
is possible to have \l\ \a\ ($c=2i{\bf A}$) even in this case,
still preserving general covariance ($b=1$) at the expense
of the trace \a\ ,
although contrary is usually preferred but not necessary.
Comparing (3.4) to (2.16) confirms that the \l\ \a\ is a genuine chiral
effect
which cannot be disposed of  by any choice of parameter $c$ in $(1,0)$
super-gravity.

It would be instructive to establish relation between the two models considered
in this work. In the non-super-symmetric case it is possible to shift from
the chiral ($\beta=\pm 1)$ and the non-chiral ($\beta=0)$ models
by introduction of a parameter in the chiral projector
$\displaystyle{P_\beta\equiv {1\over 2}\left(1+\beta\gamma^5\right)}$
and both models can be treated on the same footing [\Anoi]. In the
super-symmetric case it is not possible to obtain such a simple
simultaneous description
of both models due to the different structure of the \sfs\ in
$(1,0)$ and $(1,1)$ super-gravity, so one has to treat the two
models separately as we did. However, we can still establish relations between
these models.
In order to do so we shall use decomposition of $(1,1)$ into
$(1,0)$ \sf\ which is the analogue of the usual projection of \sf\ components.
The essence of such an approach is to eliminate
super-symmetric parameter corresponding to additional super-symmetry (in this
case $\theta_R$) by appropriate gauge choice of the super-covariant
derivative [\Ai]. In this process new $(1,0)$ super-fields
appear corresponding to ``~matter~'' gravitino and its trace of the
reduced super-symmetry.
Linearized decomposition of relevant prepotential
\sfs\ are:
$$\eqalign{
H_L^{(1,1)-}\Big\vert_{\theta^R=0}&=H_L^{(1,0)-}
\cr
D_RH_L^{(1,1)-}\Big\vert_{\theta^R=0}&=-2i\Psi_L^{(1,0)R}
\cr
H_R^{(1,1)+}\Big\vert_{\theta^R=0}&=0
\cr
D_R H_R^{(1,1)+}\Big\vert_{\theta^R=0}&=-i H_-^{(1,0)+}
\cr
H_-^{(1,1)-}\Big\vert_{\theta^R=0}&=H_-^{(1,0)-}
\cr
D_RH_-^{(1,1)-}\Big\vert_{\theta^R=0}&=2i\Psi_-^{(1,0)R}
\cr
H_L^{(1,1)L}\Big\vert_{\theta^R=0}&=H_L^{(1,0)L}
\cr
D_RH_L^{(1,1)L}\Big\vert_{\theta^R=0}&=0
\cr}
\eqno(3.5)
$$
In order to decompose $(1,1)$ effective action $(3.1)$ in terms
of $(1,0)$ \sfs\ we need decompositions of the Ricci scalar which is given
in terms of prepotentials by
$$
\bar R^{(1,1)}=i\left(\d_+D_L H_R^{(1,1)+}-\d_-D_R H_L^{(1,1)-}\right)+2ib
D_LD_RS^{(1,1)}\ .
\eqno(3.6)
$$
and, with the help of $(3.5)$ its decomposition is
$$\eqalign{
\bar R^{(1,1)}\Big\vert_{\theta^R=0}&=-2(\d_-\Psi^{(1,0)R}_L-
bD_L\Psi_-^{(1,0)R})
\cr
D_R \bar R^{(1,1)}\Big\vert_{\theta^R=0}&=\bar\Sigma^{(1,0)L}\ .
\cr}
\eqno(3.7)
$$
With the above projections one finds decomposition of $(1,1)$ effective
action $(3.1)$ as
$$\eqalign{
\ieff^{(1,1)}=\int d^2x\, d\theta^L\Bigl[&
{\bf A}\bar\Sigma^{(1,0)L}\,{1\over\nabla^2}\,D_L\bar\Sigma^{(1,0)L}+
4i{\bf A}\Psi^{(1,0)R}_L
\,{1\over\nabla^2}\,\d_-^3D_L\Psi^{(1,0)R}_L+\cr
&4b{\bf A}\Psi^{(1,0)R}_-\left(bD_L\Psi^{(1,0)R}_- -2\partial_-
\Psi^{(1,0)R}_L\right)+(c-2i{\bf A})\bar\omega_L\bar\omega_-\Bigr] \ .
\cr}
\eqno(3.8)
$$
Comparing $(3.8)$ with $(2.13)$ shows that $(3.8)$ is the effective action of
$(1,0)$
super-gravity in the case ${\bf A}=i{\bf B}$, plus additional terms due to
the ``~matter~'' gravitino and its trace.
These additional terms are there because the
effective action $(3.8)$ differs from $(2.13)$ by the fact that,
although written
in terms of $(1,0)$ \sfs, it actually has larger $(1,1)$ super-symmetry.
The \a\ equations (3.4) are decomposed as
$$\eqalign{
i\partial_-J^{(1,1)}_{+L}\Big\vert_{\theta^R=0}+{1\over 2}\partial_-D_R
J^{(1,1)}\Big\vert_{\theta^R=0}&=
(1-b)\left[2i{\bf A}\partial_+\bar\Sigma^{(1,0)L}-
(c-2i{\bf A})\,\nabla^2\,\bar\omega^{(1,0)}_L\right]
\cr
D_LD_R J^{(1,1)}_{-R}\Big\vert_{\theta^R=0}-\partial_-D_R
\tilde J^{(1,1)}\Big\vert_{\theta^R=0}&=
(b-1)\left[2i{\bf A}\partial_-\bar\Sigma^{(1,0)L}+
(c-2i{\bf A})\partial_-D_L\bar\omega^{(1,0)}_-\right]
\cr
{1\over 2}D_R J^{(1,1)}\Big\vert_{\theta^R=0}+
D_R \tilde J^{(1,1)}\Big\vert_{\theta^R=0}&=b(c+2i{\bf A})\bar\Sigma^{(1,0)L}
\cr
{1\over 2}D_R J^{(1,1)}\Big\vert_{\theta^R=0}-
D_R \tilde J^{(1,1)}\Big\vert_{\theta^R=0}&=b(c-2i{\bf A})\left(
\bar\Sigma^{(1,0)L}+2\partial_-\bar\omega^{(1,0)}_L\right)
\cr}
\eqno(3.9)
$$
plus additional \a\ equations corresponding to the $(1,0)$ currents coupling
to the matter gravitino and its trace in the effective action
$$\eqalign{
D_L J^{(1,1)}_{-R}\Big\vert_{\theta^R=0}-\partial_-
\tilde J^{(1,1)}\Big\vert_{\theta^R=0}&=
4i{\bf A}(b-1)\left[\partial_-^2\Psi_L^{(1,0)R}-b\partial_-
D_L\Psi_-^{(1,0)R}\right]
\cr
\tilde J^{(1,1)}\Big\vert_{\theta^R=0}&=
2i{\bf A}b\left[\partial_-\Psi_L^{(1,0)R}-bD_L\Psi_-^{(1,0)R}\right]
\cr}
\eqno(3.10)
$$
Relation to (3.4) is obtained through the following identification of
super-current \sfs\
$$
\eqalign{
iJ^{(1,1)}_{+L}\Big\vert_{\theta^R=0}&=J^{(1,0)}_{+L}
\cr
D_R J^{(1,1)}_{-R}\Big\vert_{\theta^R=0}&=-J^{(1,0)}_{--}
\cr
D_R J^{(1,1)}\Big\vert_{\theta^R=0}&=J^{(1,0)}
\cr
D_R \tilde J^{(1,1)}\Big\vert_{\theta^R=0}&=\tilde J^{(1,0)}
\cr}
\eqno(3.11)
$$

Further comment is needed in order to explain the presence of only one
parameter ${\bf A}$ in (3.1) and (3.8).
Prepotentials $H_L^{(1,1)-}$ and $H_R^{(1,1)+}$ appear symmetrically
and one would expect the coefficients of the corresponding non-local pieces
in the $(1,1)$
effective action to be the same. We check this point by considering
the coupling of matter to $(1,1)$ super-gravity [\Ag],
$$
I_{\rm matt.}^{(1,1)}=\int d^2x\, d\theta^L d\theta^R
(E^{(1,1)}){}^{-1}\nabla_L\Phi^{(1,1)}\nabla_R\Phi^{(1,1)}
\eqno(3.12)
$$
from which we extract linearized interactions as
$$
\eqalign{
&H_L^{(1,1)-}\d_-\Phi^{(1,1)}D_R\Phi^{(1,1)}\cr
&H_R^{(1,1)+}\d_+\Phi^{(1,1)}D_L\Phi^{(1,1)}\ .\cr}
\eqno(3.13)
$$
Projecting out component couplings (by applying $D_L$ and $D_R$ super-symmetric
derivatives to $(17)$) one finds the same contributions to non-local pieces
of the effective action (${\bf A}=i{\bf B}=3/2$).
In order to get the appropriate $(1,0)$ matter couplings we decompose
(3.13) with the help of
$$
\Phi^{(1,1)}\Big\vert_{\theta^R=0}=\Phi^{(1,0)}\ ,\quad\hbox{and}\quad
D_R\Phi^{(1,1)}\Big\vert_{\theta^R=0}=\lambda_R^{(1,0)}\ ,
\eqno(3.14)
$$
and obtain
$$
\eqalign{
&H_L^{(1,0)-}\left[\left(\d_-\Phi^{(1,0)}\right)^2+i\lambda_R\d_-
\lambda_R^{(1,0)}\right]\cr
&H_-^{(1,0)+}\d_+\Phi^{(1,0)} D_L\Phi^{(1,0)}\cr}
\eqno(3.15)
$$

Now, we can understand the possible absence of \l\ \a\ in this
case since the above projection introduces extra right-handed fermion
\sf\ in addition to the scalar \sf\ described in Sect.(2). This
additional contribution, whose presence is due to the larger $(1,1)$
super-symmetry, compensate for the missing chirality.

\chapter{Summary}

We have described the super-gravitational, Weyl and \l\ \as, on the same,
most general footing, both in $(1,0)$(chiral) and $(1,1)$(non-chiral)
super-gravity. It is shown that, in order to do so, we need {\it two}
arbitrary regularization dependent parameters similarly to the situation
already found in the non-super-symmetric gravity model [\Anoi].
Super-symmetry introduces an additional freedom represented
by the parameters in the
{\it non-local} piece of the effective action. This occurs because matter \sfs\
contain more ingredients which are not present in
non-super-symmetric models. Nevertheless,
$(1,0)$ chiral model has in general genuine \l\ \a.

In this work we have
considered only super-gravitational \as, however it would be interesting
to include super-Schwinger model \as\ as well. Preliminary investigation
of the super-Schwinger model has been done in the case of $(1,1)$
super-symmetry [\Ajose] showing the same arbitrariness in terms of
a single parameter as in the non-super-symmetric case, together with some
new features related to the breakdown of the duality relation between
axial and vector gauge currents. From this point of view,
it would particularly interesting to investigate
$(2,2)$ super-gravity where the axial gauge field is a member of the
gravitational super-multiplet. Therefore  one would expect
the same parameter to
describe both gravitational and gauge \as.  Decomposing the
gravitational super-multiplet to lower
super-symmetry realizations would give new insight into these models
since, so far, Schwinger and gravitational \as\ have been treated
separately in terms of independent parameters. This problem is now under
investigation.

\refout
\bye